\documentclass{desyproc}

\begin{document}
\title{WISPDMX: A haloscope for WISP Dark Matter between 0.8-2 $\mu$eV}

\author{{\slshape  Le Hoang Nguyen$^1$, Dieter Horns$^1$, Andrei Lobanov$^{1,}$$^3$, Andreas Ringwald $^2$,}\\
$^1$: Institut f{\"u}r Experimentalphysik, Universit{\"a}t Hamburg, Germany. \\
$^2$: Deutsches Elektronen-Synchrotron (DESY), Hamburg, Germany.\\
$^3$: Max-Planck-Institut f{\"u}r Radioastronomie, Bonn, Germany.}

\contribID{Nguyen Le Hoang}

\confID{11832}  
\desyproc{DESY-PROC-2015-02}
\acronym{Patras 2015} 
\doi  

\maketitle

\begin{abstract}
Weakly interactive slim particles (WISPs), including the QCD axion, axion-like particles (ALPs), and hidden photons, are considered to be
strong candidates for the dark matter carrier particle.  The microwave cavity experiment WISPDMX is the first direct WISP dark matter search experiment probing the particle masses in the 0.8-2.0 $\mu$eV range. The first stage of WISPDMX measurements has been completed at nominal resonant frequencies of the cavity. The second stage of WISPDMX is presently being prepared, targeting hidden photons and axions within 60\% of the entire 0.8-2.0 $\mu$eV mass range.
\end{abstract}
\section{Introduction}

\begin{figure}[h!]
 \centering 
 \includegraphics[width=15cm]{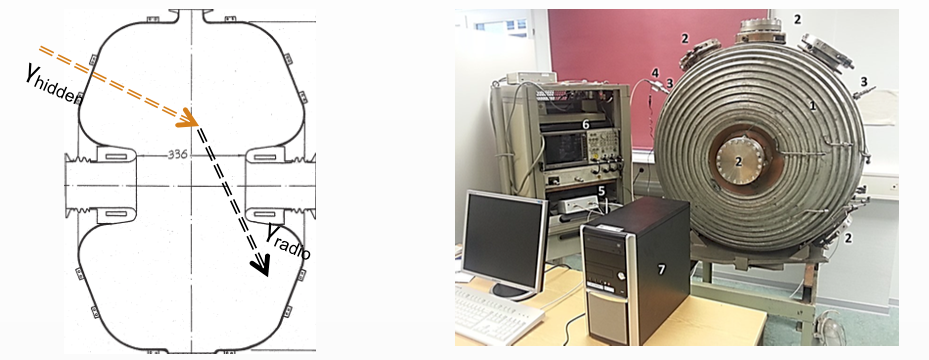}
 \caption{WISPDMX utilises a 208-MHz resonant cavity used at the DESY HERA accelerator and plans to make use of the H1 solenoid magnet. The cavity has a resonant amplification factor Q = 46000 at the ground$\text{TM}_{010}$ mode. The figures showing simple sketches of the 208-MHz resonant cavity with possible conversion from HP to RF radiation (Left), and the first stage's experiment setup of WISPDMX.}
 \label{fig1}
\end{figure}

Weakly Interacting Slim (Sub-eV) Particles \cite{ref1,ref2,Arias:2012az} are promising candidates for a dark matter particle and together with WIMPs, axions, and hidden photons are an attractive field for DM searches. The most favoured particle mass range for axion dark matter is between $10^{-7}$ and $10^{-3}$ eV which makes radio measurement at frequencies below 240 GHz a prime experimental tool for axion detection. Searches for the WISPs DM are cataloged into three types: purely laboratory experiments (Light-Shining Through Walls Experiments) using optical photons, helioscopes observing WISPs emitted by the Sun, and haloscopes which are searching for dark matter constituents. \\
WISP Dark Matter eXperiment (WISPDMX) has been initiated at DESY and the University of Hamburg \cite{ref3}, aiming at covering the 0.8- 2 $\mu$eV mass range, probing into the DM-favored coupling strengths. WISPDMX has three phases. Phase I: hidden photon searches at nominal resonances of the cavity, Phase II: cavity tuning for searches and Phase III: Axion Like Particles searches with the adaption of HERA magnet.\\
The experiment utilises a 208-MHz resonant cavity (Fig. \ref{fig1}) used at the DESY HERA accelerator and plans to make use of the H1 solenoid magnet \cite{accelerator}. The cavity has a volume of 460 litres and a resonant amplification factor Q = 46000 at the ground $\text{TM}_{010}$ mode. The H1 magnet provides B = 1.15T in a volume of 7.2 m$^3$. The signal is amplified by a broad-band 0.2-1GHz amplifier with a system temperature of 100K. Broad-band digitisation and FFT analysis of the signal are performed using a commercial 12-bit spectral analyser, enabling simultaneous measurements at several resonant modes. The cavity tuning can be provided with the use of a plunger assembly inserted into the cavity. The original plunger assembly used with the HERA cavity for the accelerator needs to be modified for the tuning. \\
 
\section{WISPDMX status.}

\subsection{Result from Phase I}
\begin{figure}[h!]  
 \centering
 \includegraphics[width=7cm]{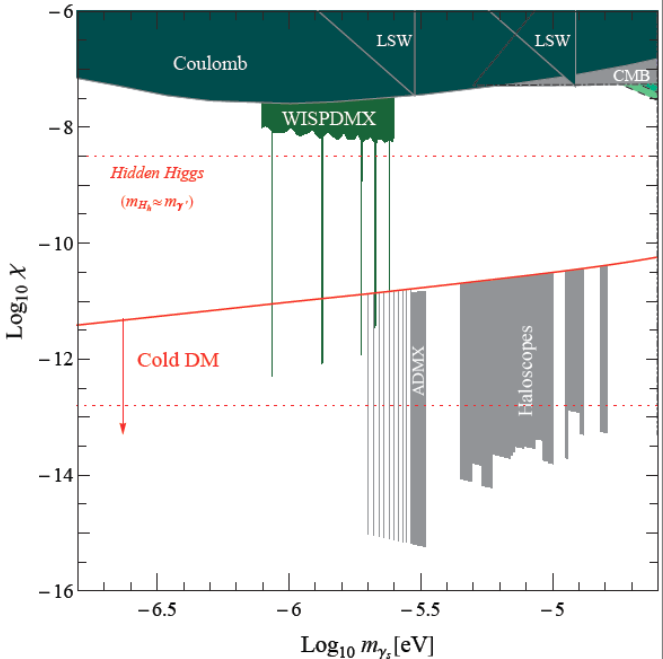}
 \caption{Current exclusion limit set by Phase I of WISPDMX \cite{ref4}: evaluating the broadband signal (600 MHz) under 40.3 dB amplification. The useful range is 180-600 MHz and frequency resolution of $\Delta \nu$ = 572 Hz. The turquoise colour lines are exclusion limit set by ADMX.}
 \label{fig2}
\end{figure}

\begin{figure}[h!] \centering 
 \includegraphics[width=13cm]{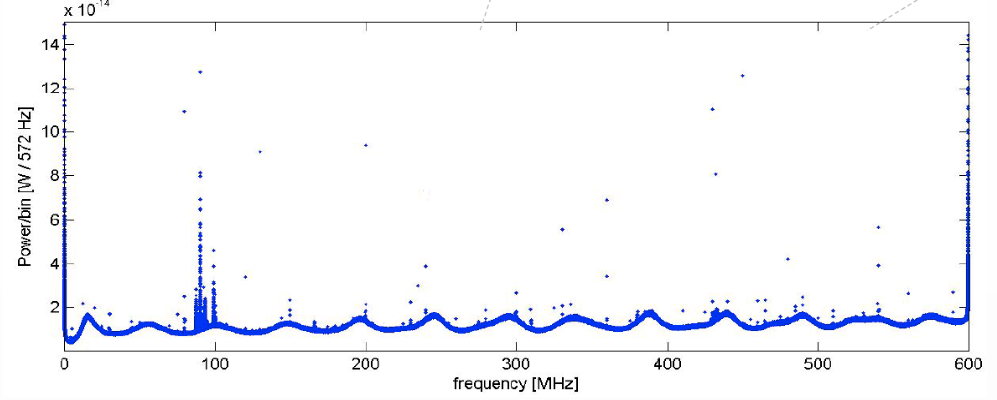}
  \caption{Broadband noise spectrum obtains from the Phase I of WISPDMX.}
 \label{fig3}
\end{figure}

In Phase I, we evaluated the broadband signal, by using the commercial ADC card (1.8 MSPS and 12 bits), measurements at the nominal frequencies at the resonant modes setting up the initial exclusion limits and obtaining the noise spectrum shown in figure \ref{fig2}, \ref{fig3}.  \\

\subsection{Phase II: Development and Preliminary Result.}
\subsubsection{Phase II Experiment Setup.}
\begin{figure}[h!]
 \centering 
 \includegraphics[width=9cm]{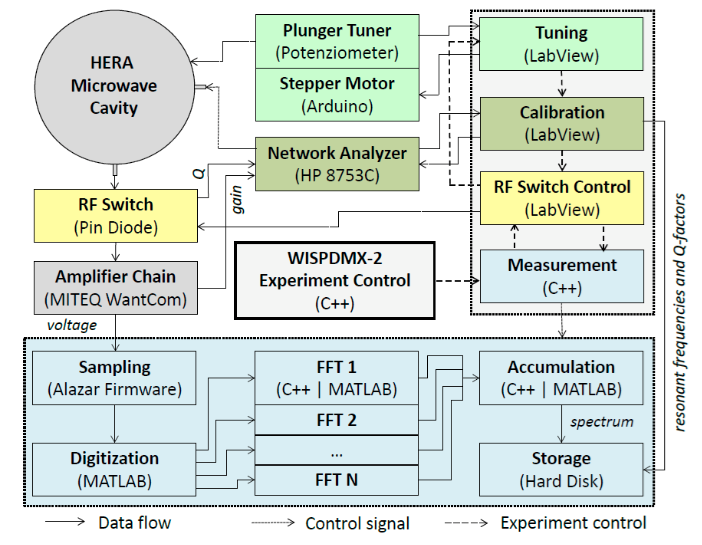}
 \caption{The flowchart illustrating the measurement procedures designed for Phase II of the WISPDMX. The tunning is made with a plunger driven by a stepper motor. The frequency calibration is performed with help of a network analyser. The signal is amplified, digitised and analysed with a commercial digitised control by Matlab/C++ software. The overall experiment control is set within the Labview environment.  }
 \label{fig4}
\end{figure}

In Phase II, we plan to perform simultaneous multiple mode measurements (with frequency calibration and broad band signal recording) with the help of tuning plungers. We enhance the experiment with automatic tuning, continuous calibration and signal recording (Fig. \ref{fig4}). \\

 The tuning plunger plays an important role in Phase II in searching for WISPs over a broad mass-range. The plunger assembly should provide effective coverage up to 56\% of the 200-500 MHz range. The first plunger has been designed and manufactured, the second one is under construction. The tuning will be accomplished with a plunger assembly providing a 2 MHz tuning range of the ground mode and up to 30 MHz for the higher modes. 
\subsubsection{Phase II Preliminary Result.}
\begin{figure}[h!]
 \centering
  \includegraphics[width=10cm]{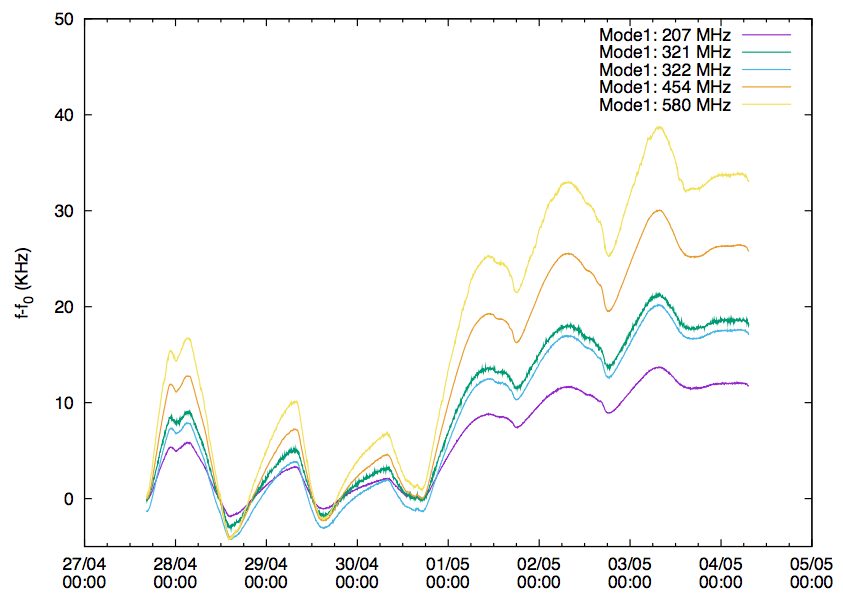}
  \caption{The shifting of 5 resonant modes due to the environment temperature. The frequency shift with respect to the temperature is 3KHz/K.}
   \label{fig5}
\end{figure}

We study the reaction of the cavity to the temperature and atmospheric pressure changes by measuring the resonant modes of the cavity, and studying their dependence on changes with both of these quantities (see Fig. \ref{fig5}). This study has yielded a good calibration that can be implemented into measurement with respect to the variability of temperature and atmospheric pressure. 

\section{Conclusion}
The WISPDMX component’s and tools are 60\% completed for the Phase II, with the second plungers for the cavity to be manufactured before the end of 2015 and software development to be ready for a preliminary run with one plunger. Further tests on the  frequency calibration will be made in order to ensure frequency fidelity and accuracy at the desired spectral sensitivity in and out at the resonance.

\end{document}